\documentclass[twocolumn]{aastex63}

\usepackage{physics, xcolor}
\usepackage{comment}
\newcommand{\pc}{\mathrm{pc}}
\newcommand{\mo}{\mathrm{M}_\odot}

\shorttitle{On the Correlation between Hot Jupiters and Stellar Clustering}

\begin{document}

\title{On the Correlation between Hot Jupiters and Stellar Clustering: High-eccentricity Migration Induced by Stellar Flybys}

\correspondingauthor{Laetitia Rodet}
\email{lbr63@cornell.edu}

\author[0000-0002-1259-3312]{Laetitia Rodet}
\author[0000-0001-8283-3425]{Yubo Su}
\affiliation{Cornell Center for Astrophysics and Planetary Science, Department of Astronomy, Cornell University, Ithaca, NY 14853, USA}
\author[0000-0002-1934-6250]{Dong Lai}
\affiliation{Cornell Center for Astrophysics and Planetary Science, Department of Astronomy, Cornell University, Ithaca, NY 14853, USA} 



\begin{abstract}

A recent observational study suggests that the occurrence of hot Jupiters (HJs) around solar-type stars is correlated with stellar clustering. We study a new scenario for HJ formation, called ``Flyby Induced High-e Migration'', that may help explain this correlation. In this scenario, stellar flybys excite the eccentricity and inclination of an outer companion (giant planet, brown dwarf, or low-mass star) at large distance (10--300 au), which then triggers high-e migration of an inner cold Jupiter (at a few astronomical units) through the combined effects of von Zeipel--Lidov--Kozai (ZLK) eccentricity oscillation and tidal dissipation. Using semianalytical calculations of the effective ZLK inclination window, together with numerical simulations of stellar flybys, we obtain the analytic estimate for the HJ occurrence rate in this formation scenario. We find that this ``flyby induced high-e migration'' could account for a significant fraction of the observed HJ population, although the result depends on several uncertain parameters, including the density and lifetime of birth stellar clusters, and the occurrence rate of the ``cold Jupiter + outer companion'' systems.

\end{abstract}



\section{Introduction} \label{sec:intro}


Hot Jupiters (HJs), giant planets orbiting with very short periods ($\sim$ 3 days) are found around 1\% of FGK stars \citep[see][and references therein]{dawson_origins_2018}. HJs can form in three ways: in situ, through disk-driven migration, or through high-eccentricity migration. There is currently no consensus on the predominant channel for their origin.



Recent work by \cite{winter_stellar_2020} has revealed an intriguing correlation between the occurrence of HJs and stellar clustering. For each exoplanet-hosting star in their sample, the authors computed the local stellar phase-space density of the star and its neighbors (within 40 pc) using Gaia DR2 \citep{gaia_collaboration_gaia_2018}. They determined whether the exoplanet host was in a relatively low or high stellar density zone compared to its neighbours, and concluded that HJs were preferably found in local stellar phase-space overdensities. The origin of this correlation is puzzling, as stellar clustering was thought to affect mostly the outer part of planetary systems in very dense environments \citep{laughlin_modification_1998,malmberg_effects_2011,parker_effects_2012,cai_stability_2017,li_flyby_2020}.

The formation of HJs in situ or through disk-driven migration is not directly correlated to the stellar environment. However, high-e migration may correlate with stellar density. This mechanism relies on the excitation of a cold Jupiter's eccentricity by an outer perturber. The semimajor axis of the planet is then decreased by tidal dissipation at each periastron passage \citep[e.g.,][]{wu_planet_2003,fabrycky_shrinking_2007,nagasawa_formation_2008,wu_secular_2011,beauge_multiple-planet_2012,naoz_formation_2012,petrovich_hot_2015,petrovich_steady-state_2015,anderson_formation_2016,teyssandier_formation_2019,vick_chaotic_2019}. The occurrence rate of this formation path may depend on the frequency of stellar flybys, through dynamical interactions between the passing stars and the outer planetary system \citep[e.g.,][]{shara_dynamical_2016}. In \cite{wang_hot_2020}, the authors considered a scenario where a stellar flyby excites an outer Saturn, which in turn raises the eccentricity of the inner Jupiter by planet-planet scattering or through the von Zeipel--Lidov--Kozai \citep[ZLK,][]{von_zeipel_sur_1910,kozai_secular_1962,lidov_evolution_1962} mechanism. They estimated numerically the resulting occurrence rate of HJs in a virialized cluster, and showed that this formation path has a negligible rate. However, this study is restricted to specific ranges of cluster parameters and initial planetary systems, and the numerical approach limits the generality of the results.


In this paper, we examine the likelihood of HJs forming through high-e migration triggered by flybys in a stellar environment, using a combination of analytical calculations (for high-e migration) and numerical simulations (for stellar flybys). In Section~\ref{sec:scenario} we outline the proposed scenario and its key ingredients. In Section~\ref{sec:kozai}, we review the conditions for the ZLK mechanism to produce an HJ, and extend previous works to examine the case of a planetary-mass perturber. In Section~\ref{sec:flyby}, we evaluate the extent to which a stellar encounter can raise the eccentricity and inclination of an outer companion. In Section~\ref{sec:rate}, we derive the occurrence rate of HJs as a function of the properties of the system and its stellar neighborhood. Finally, in Section~\ref{sec:conclusion}, we conclude and discuss alternative explanations for the observed correlations between HJs and stellar overdensities.


\section{Flyby Induced High-e Migration Scenario}
\label{sec:scenario}

According to the current understanding of planetary formation, giant planets form preferentially beyond the snow line at a few astronomical units. Their migration to the close neighborhoods of their host stars can be triggered by a combination of eccentricity excitation and tidal dissipation---the so-called high-e migration mechanism (see references in Section~\ref{sec:intro}). This requires the presence of a misaligned and eccentric outer companion (see Section~\ref{sec:kozai}). To evaluate the probability to form HJs through high-e migration, we thus need to estimate the occurrence rate of such companions.

On the one hand, misaligned companions could be naturally associated with a stellar binary. The orientation of stellar companions with respect to the protoplanetary disk plane of the primary is expected to be random at large separations. This is also likely the case for at least some fraction of the substellar companions (brown dwarfs). This possibility is examined in Section~\ref{sec:conclusion}.

On the other hand, a scattering encounter between a planetary system and a passing star could raise the eccentricity and inclination of an initially coplanar and circular outer companion, which then drives the inner cold Jupiter into a high-e orbit, leading to HJ formation. The probability of this HJ formation channel is strongly dependent on the stellar density. The combination of flyby and high-e migration could account for the observed correlation between HJ occurrence and stellar overdensity. High-e migration can be triggered by the ZLK mechanism or by planet-planet scattering. This paper will focus on the former, and the latter is discussed in Section~\ref{sec:conclusion}.




For the basic setup, we consider two bodies around a solar-type star (with mass $M_1$), an inner cold Jupiter ($m_{\rm J}$) at $a_{\rm J} \sim$ 5 au and an outer companion/perturber ($m_{\rm p}$) at larger separations ($a_{\rm p} \in 10$--500 au). We assume that the initial orbits of $m_{\rm J}$ and $m_{\rm p}$ are coplanar and circular, as is expected from their formation in a gaseous protoplanetary disk (note that the perturber could also be a brown dwarf or low-mass star). We consider a passing star ($M_2$) with velocity at infinity $v_\infty$ and periastron $q$. In the following sections, we will examine the requirements for the passing star to trigger the high-e migration of the cold Jupiter via excitation of the outer companion's orbit.


\section{Effective range of ZLK eccentricity excitation and High-e Migration}
\label{sec:kozai}

\subsection{Semimajor Axis Window}

Not every semimajor axis ratio $a_{\rm J}/a_{\rm p}$ can lead to the formation of HJs. High-e migration requires the inner planet ($m_{\rm J}$) to attain a sufficiently large eccentricity. The maximum eccentricity depends on $i_{\rm p}$, the mutual inclination between the planet and the companion ($m_{\rm p}$), and the competition between eccentricity driving by the ZLK mechanism and its suppression due to short-range forces \citep[e.g.,][hereafter LML15]{fabrycky_shrinking_2007,liu_suppression_2015}. In our case, the most important of these forces is the tidal force from the star on the inner planet. Over all possible values of $i_{\rm p}$, the inner planet cannot become more eccentric than the limiting eccentricity $e_\mathrm{lim}$, as determined by LML15:
\begin{equation}
    \frac{\dot\omega_\mathrm{tides}}{\dot\omega_\mathrm{ZLK}} \big|_{e_{\rm J}=e_\mathrm{lim}}=\frac{81}{8},
\end{equation}
where $\dot\omega_\mathrm{tides}$ and $\dot\omega_\mathrm{ZLK}$ are the precession frequencies due to the tidal force and ZLK mechanism, respectively. They are given by:
\begin{eqnarray}
	\dot\omega_\mathrm{tides}={}&&\frac{15}{2} k_{2,{\rm J}} \frac{M_1}{m_{\rm J}} \left(\frac{R_{\rm J}}{a_{\rm J}}\right)^5 \frac{1+\frac{3}{2}e_{\rm J}^2 + \frac{1}{8}e_{\rm J}^4}{(1-e_{\rm J}^2)^5} n_{\rm J},\\
	\dot\omega_\mathrm{ZLK}={}&& \frac{m_{\rm p}}{M_1} \left(\frac{a_{\rm J}}{a_{\rm p,eff}}\right)^3 \frac{n_{\rm J}}{\sqrt{1-e_{\rm J}^2}},
\end{eqnarray}
where $a_{\rm p,eff} = a_{\rm p} \sqrt{1-e_{\rm p}^2}$, $k_{2,{\rm J}}$ is the tidal Love number of the planet, $R_{\rm J}$ is its radius, and $n_{\rm J} \equiv \sqrt{G M_1 / a_{\rm J}^3}$ is its mean motion. We find
\begin{eqnarray}
	1-e_{\mathrm{lim}} \approx{}&& 10^{-3} \left(\frac{k_{2,{\rm J}}}{0.37}\right)^\frac{2}{9} \left(\frac{R_{\rm J}}{1~ \mathrm{R_{Jup}}}\right)^\frac{10}{9} \left(\frac{a_{\rm p,eff}}{40~\mathrm{au}}\right)^\frac{2}{3}  \nonumber\\	&& \times \left(\frac{m_{\rm p}}{5~\mathrm{M_{Jup}}}\right)^{-\frac{2}{9}}
	\left(\frac{m_{\rm J}}{1~\mathrm{M_{Jup}}}\right)^{-\frac{2}{9}}\nonumber\\ && \times  \left(\frac{M_1}{1~\mathrm{M_\odot}}\right)^\frac{4}{9} \left(\frac{a_{\rm J}}{5~\mathrm{au}}\right)^{-\frac{16}{9}}.\label{eq:elim}
\end{eqnarray}
To produce an HJ with a (circular) semimajor axis $a_{\rm HJ}$ ($\sim 0.04$ au, corresponding to a 3 day orbit), we require the pericenter distance $a_{\rm J} (1-e_{\rm J})$ to reach below $a_{\rm HJ}/2$. This gives
\begin{eqnarray}
	&&\frac{a_{\rm J}}{a_{\rm p, eff}}\gtrsim \frac{1}{50} \left(\frac{k_{2, {\rm J}}}{0.37}\right)^\frac{2}{7} \left(\frac{R_{\rm J}}{1~ \mathrm{R_{Jup}}}\right)^\frac{10}{7} \left(\frac{m_{\rm J}}{1~\mathrm{M_{Jup}}}\right)^{-\frac{2}{7}} \nonumber\\
	&&
	{}\times \left(\frac{M_1}{1~\mathrm{M_\odot}}\right)^\frac{4}{7} \left(\frac{m_{\rm p}}{5~\mathrm{M_{Jup}}}\right)^{-\frac{2}{7}}  \left(\frac{a_{\rm p, eff}}{40~\mathrm{au}}\right)^{-\frac{1}{7}} \left(\frac{a_\mathrm{HJ}}{0.04~\mathrm{au}}\right)^{-\frac{9}{7}}.\label{eq:aratio}
\end{eqnarray}
The possibility of creating an HJ thus depends mostly on the semimajor axis ratio $a_{\rm J}/a_{\rm p}$, and weakly on $a_{\rm p}$ alone. For giant planets initially at 5 au and the fiducial parameters, Equation~\eqref{eq:aratio} amounts to $a_{\rm p, eff} \lesssim 300$ au. As most planets are expected to lie inside this limit, it follows that Equation~\eqref{eq:aratio} does not provide a strong constraint on the semimajor axis of the outer companion. 

\subsection{Eccentricity and Inclination Window}

\begin{figure}
	\centering
	\includegraphics[width=\linewidth]{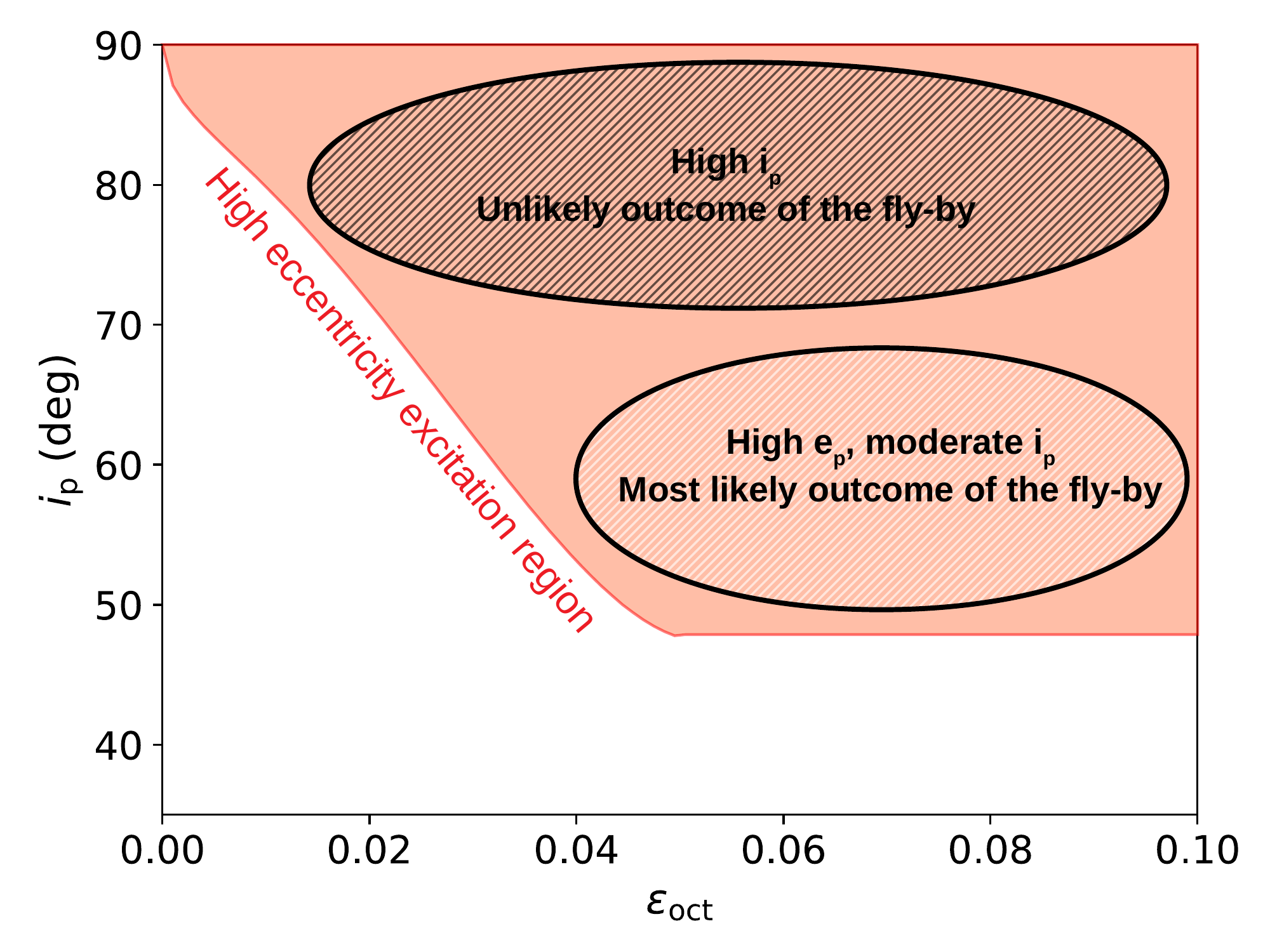}
	\caption{High-eccentricity excitation region for the inner planet in the ZLK problem at the octupole order, in the $\varepsilon_\mathrm{oct}$--$i_\mathrm{p}$ phase space (where $i_{\rm p}$ is the inclination of the external perturber, and $\varepsilon_\mathrm{oct}$ is given by Equation~\ref{eq:epsoct}). The boundary of the light red zone is given by the fitting formula of \cite{munoz_formation_2016}, and systems in this zone can attain extreme eccentricity $e_{\rm lim}$ (Equation~\eqref{eq:elim}), assuming the inner planet has a negligible angular momentum compared to that of the outer perturber. High inclinations are harder to produce with a flyby than high eccentricities (corresponding to high values of $\varepsilon_\mathrm{oct}$, see Section~\ref{sec:flyby}).}\label{fig:fig1}
\end{figure}

\begin{figure*}
    \centering
    \includegraphics[width=\linewidth,trim={1.5cm 0.5cm 2.5cm 0.5cm}]{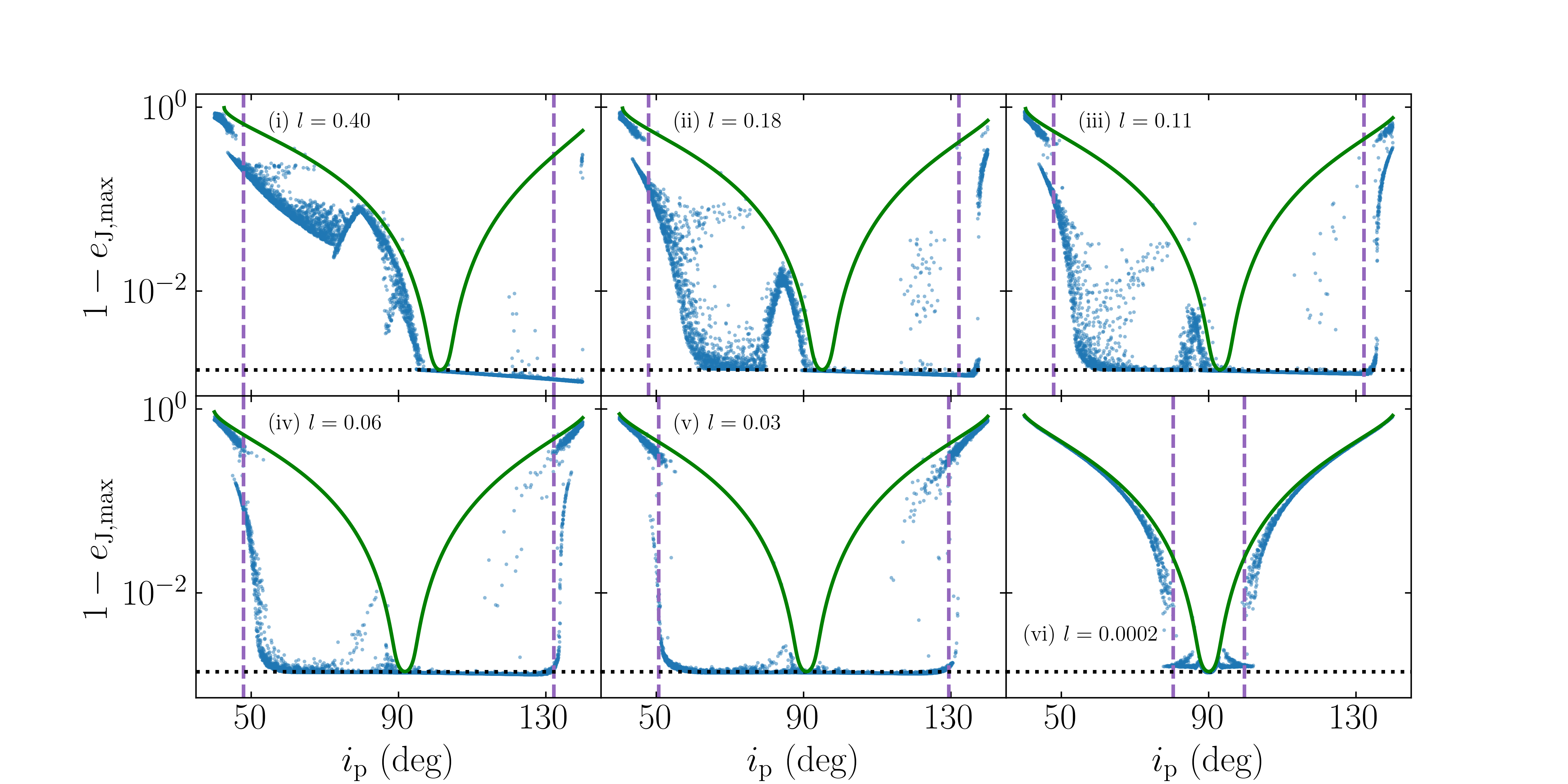}
    \caption{Maximum eccentricity of the inner planet (as $1 - e_{\rm J, \max}$) vs. initial inclination $i_{\rm p}$ for different values of the initial angular momentum ratio $l$ (Equation~\eqref{eq:angularmomentum} evaluated at $e_{\rm J} = 0$; labeled) and semimajor axis $a_{\rm p}$ such that $t_{\rm ZLK}$ (Equation~\eqref{eq:t_LK}) is held constant ($a_{\rm p} = 50$, $63$, $72$, $85$, $108$, and $500$ au respectively). The corresponding perturber masses are $m_{\rm p} = M_{\rm Jup}$, $2 M_{\rm Jup}$, $3 M_{\rm Jup}$, $5 M_{\rm Jup}$, $10 M_{\rm Jup}$, and $M_{\odot}$ respectively. In all cases, we take $e_{\rm p} = 0.6$, $M_1 = M_{\odot}$, $m_{\rm J} = M_{\rm Jup}$, and $a_{\rm J} = 5\;\mathrm{AU}$. The blue dots denote the maximum eccentricities attained by $m_{\rm J}$ when the system is integrated for $500 t_{\rm ZLK}$. The green line illustrates the analytical $e_{\rm J, \max}(i_{\rm p})$ curve when the octupole effect is neglected (see Equation~50 of LML15). The purple vertical lines denote the inclination window for extreme eccentricity excitation as given by the fitting formula of MLL16 (see Figure~\ref{fig:fig1} or Equation~7 of MLL16); for the simulated systems in the six panels, $\varepsilon_{\rm oct} = 0.09$, $0.07$, $0.06$, $0.05$, $0.04$, and $0.009$, respectively. The horizontal dashed line denotes $e_{\lim}$ as given by Equation~\eqref{eq:elim}. When $m_{\rm p} \lesssim 3M_{\rm J}$, corresponding to $l \gtrsim 0.1$, the angular momentum of the inner planet is non-negligible, and it is much more difficult for prograde outer companions ($i_{\rm p} < 90^\circ$) to excite $e_{\rm J}$ to $e_{\lim}$ than it is for retrograde outer companions ($i_{\rm p} > 90^\circ$). }\label{fig:YS_emaxes}
\end{figure*}

In the ZLK mechanism, at the quadrupole order, an initially circular planet can only reach extreme eccentricities if the outer companion's inclination is close to $90^\circ$. This picture is valid under two assumptions: the inner planet is effectively a ``test particle'' (i.e., its orbital angular momentum is negligible compared to that of the outer companion) and the octupole-order corrections are sufficiently weak. In the test-particle limit, it has been shown that the octupole-order effect can induce extreme eccentricities ($e_{\rm J} \simeq e_\mathrm{lim}$) in the inner orbit if the outer orbit's inclination belongs to a finite window around $90^\circ$ \citep[LML15;][hereafter MLL16]{munoz_formation_2016} whose size depends on the octupole parameter:
\begin{equation}
	\varepsilon_\mathrm{oct} = \frac{a_{\rm J}}{a_{\rm p}} \frac{e_{\rm p}}{1-e_{\rm p}^2}.\label{eq:epsoct}
\end{equation}
As  $\varepsilon_\mathrm{oct}$ increases, the inclination window grows until it saturates at $[48^\circ, 132^\circ]$ for $\varepsilon_\mathrm{oct} \gtrsim 0.05$ (see Figure~\ref{fig:fig1}). This ``symmetric'' inclination window is valid in the test-particle limit, when the angular momentum ratio $l$ between the inner planet and the outer companion is small, i.e.,
\begin{equation}
	l \equiv \frac{m_{\rm J}}{m_{\rm p}} \sqrt\frac{a_{\rm J} (1-e_{\rm J}^2)}{a_{\rm p} (1-e_{\rm p}^2)} \ll 1. \label{eq:angularmomentum}
\end{equation}
However, when both $l$ and $\varepsilon_{\rm oct}$ are not negligible, the inclination window obtained by MLL16 cannot predict whether the inner orbit reaches very high eccentricities. To understand the critical value of $l$ below which the prescription of MLL16 is accurate, we have carried out new simulations for the evolution of the inner and outer orbits to octupole-order, using the secular equations given in LML15. We also include general relativistic periastron advance and tidal distortion of the giant planet following LML15. We ignore the orbital decay of the inner planet due to tidal dissipation (which is expected to occur over long timescales). To isolate the impact of different values of $l$, we vary $m_{\rm p}$ and $a_{\rm p}$ such that the quadrupole order ZLK timescale, given by
\begin{equation}
    t_{\rm ZLK}^{-1} \equiv \frac{m_{\rm p}}{M_1}
        \left(\frac{a_{\rm J}}{a_{\rm p, eff}}\right)^3n_{\rm J},
        \label{eq:t_LK}
\end{equation}
is constant. In particular, we fix $e_{\rm p} = 0.6$ and the initial $e_{\rm J} = 10^{-3}$ and consider six values of $m_{\rm p} = \lbrace{1, 2, 3, 5, 10}\rbrace \times M_{\rm Jup}$ and $m_{\rm p} = M_{\odot}$, while adjusting $a_{\rm p}$ accordingly. For each value of $m_{\rm p}$, we further consider $2000$ uniformly spaced initial inclinations $i_{\rm p} \in \left[{40^\circ, 140^\circ}\right]$. Then, for each inclination, we run three simulations while randomly choosing $\Omega, \omega \in [0, 2\pi)$ for both the inner and outer orbits\footnote{Note that the dynamics of the system are essentially independent of the initial $\Omega$ and $\omega$, but the chaotic nature of the octupole-order ZLK effect causes the detailed evolution to differ for different initial angles. Thus, just three simulations with different $\Omega$ and $\omega$, in conjunction with our dense grid of $i_{\rm p}$, are enough to explore the range of behaviors for given $m_{\rm p}$ and $i_{\rm p}$.}, the longitude of the ascending node and argument of periapsis, respectively, totaling $6000$ simulations per combination of $m_{\rm p}$ and $a_{\rm p}$. We run each simulation for $500 t_{\rm ZLK}$ and measure the maximum eccentricity attained by the inner planet.
Figure~\ref{fig:YS_emaxes} depicts our numerical results. We see that the inclination window predicted by MLL16 is accurate when the initial angular momentum ratio satisfies
\begin{equation}
    l \lesssim 0.1.\label{eq:angularmomentumprecise}
\end{equation}
For $l\gtrsim 0.1$, extreme eccentricity can be achieved only for inclinations larger than the test-particle result. As we shall see in Section 4, for initially coplanar systems, very high inclinations are an unlikely outcome of flybys. Thus, in the following, we will restrict our attention to systems satisfying Equation~\eqref{eq:angularmomentumprecise}. Assuming $e_{\rm p} = 0.6$, this would require $m_{\rm p} > 4~\mathrm{M_{Jup}}$ for $a_{\rm p}/a_{\rm J} = 10$; and $m_{\rm p} > 2~\mathrm{M_{Jup}}$ for $a_{\rm p}/a_{\rm J} = 50$. Note that the limiting eccentricity $e_{\rm lim}$ depends on $l^2$ \citep[][Equation~(26)]{anderson_eccentricity_2017}, and is almost unchanged from Equation~\eqref{eq:elim} in the regimes we consider, as can be seen in Figure~\ref{fig:YS_emaxes}.


\begin{figure*}
	\centering
	\includegraphics[width=\linewidth]{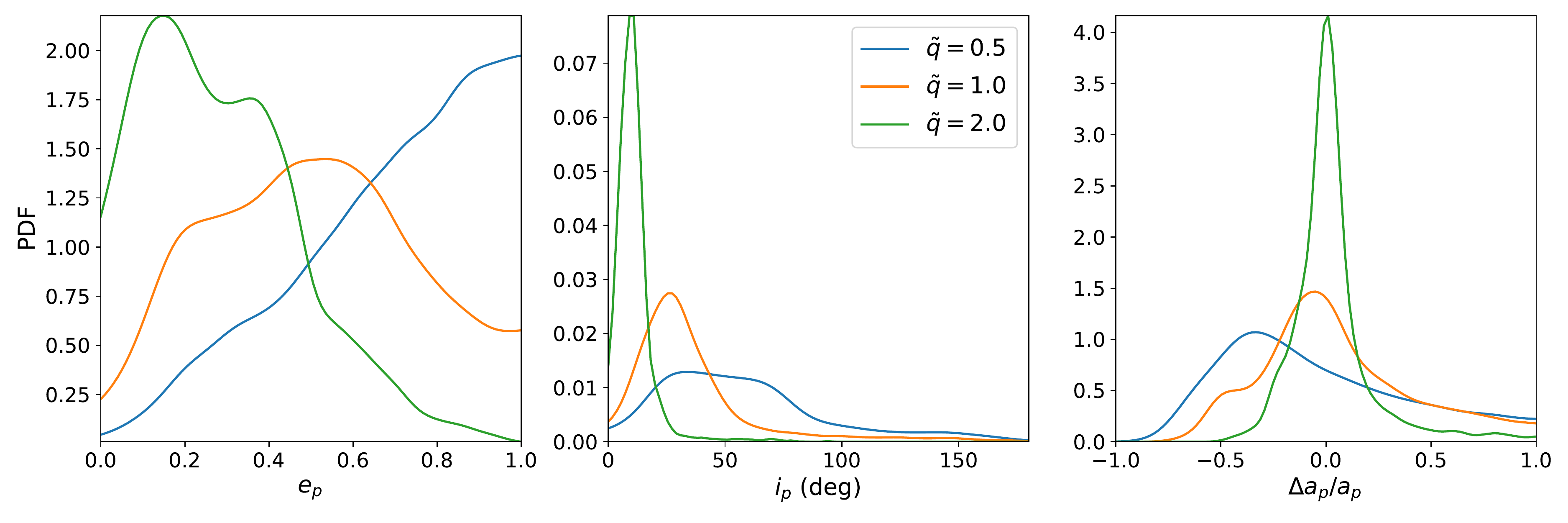}
	\caption{Probability distribution of the orbital elements (eccentricity, inclination, and fraction change of semimajor axis) of the outer companion ($m_{\rm p}$) after stellar flybys. Numerical results for three values of the pericenter distance $q$ of the passing star are shown, with $\tilde{q} \equiv q/a_{\rm p} = 0.5, 1$ and $2$. The companion is initially circular (and coplanar with the inner planet). Only the configurations where the companion remains bound are plotted here.}\label{fig:several_q}
\end{figure*}

\section{Effect of flyby}
\label{sec:flyby}

The goal of this section is to estimate the percentage of close encounters that can drive the outer companion/perturber into the ZLK window---which we will refer to as ``successful flybys''. Recall that in our scenario the outer body ($m_{\rm p}$) serves as a perturber that drives the inner planet into a high-e orbit. A successful flyby should raise significantly both the octupole parameter $\varepsilon_\mathrm{oct}$ (or eccentricity) and the inclination of the outer body (Figure~\ref{fig:fig1}). In the following, we will show that this is roughly equivalent to raising the inclination $i_{\rm p}$ to at least $48^\circ$.

The impact of a flyby on the outer companion/perturber depends on the dimensionless distance at closest approach $\tilde{q} \equiv q/a_{\rm p}$, where $q$ is the flyby periastron. For large $\tilde{q}$, we can average over time the orbit of the companion and the trajectory of the passing star and analytically compute the final inclination $i_{\rm p}$ and eccentricity $e_{\rm p}$ of the orbit of the outer companion \citep{heggie_effect_1996,rodet_odea_2019}. These analytical expressions hold only for $q \gtrsim 3a_{\rm p}$, and the maximum inclination increase is about $6^\circ$, not enough for the system to enter the ZLK extreme eccentricity excitation window (which requires at least a misalignment of $48^\circ$). The impact of a flyby with a smaller periastron is chaotic and can only be studied numerically. We thus conduct $N$-body simulations with the \textsc{ias15} integrator of the \textsc{rebound} package \citep{rein_rebound_2012,rein_ias15_2015}. In order to limit the number of parameters, the simulations include two equal-mass stars $M_1 = M_2$ and a test particle (representing the low-mass companion $m_{\rm p}$). The integration time is chosen so that the distance between $M_1$ and $M_2$ is equal to $100a_\mathrm{p}$ at the beginning and end of the simulations, and the time-step is adaptive.

For stars $M_1$ and $M_2$ approaching each other with relative velocity $v_\infty$ \citep[$\sim 1$ km s$^{-1}$, typical of young clusters; see][]{wright_kinematics_2018}, the velocity at periastron $q$ is given by
\begin{eqnarray}
	v_\mathrm{peri}^2={} && v_\infty^2 + \frac{2GM_\mathrm{tot}}{q}\nonumber\\
	{}={}&& v_\infty^2 + (10~\mathrm{km/s})^2 \left(\frac{50~\mathrm{au}}{q}\right)\left(\frac{M_\mathrm{tot}}{2~M_\odot}\right),
\end{eqnarray}
where $M_\mathrm{tot} = M_1 + M_2$. For the regime we are considering, the gravitational focusing term dominates and $v_\infty$ is negligible. This translates to a stellar eccentricity close to 1.
In our simulations, we adopt $e=1.1$ for all encounters. We have checked that the impacts of the flybys did not vary significantly with $e$ as long as it remains close to $1$.

\begin{figure*}
	\centering
	\includegraphics[width=\linewidth]{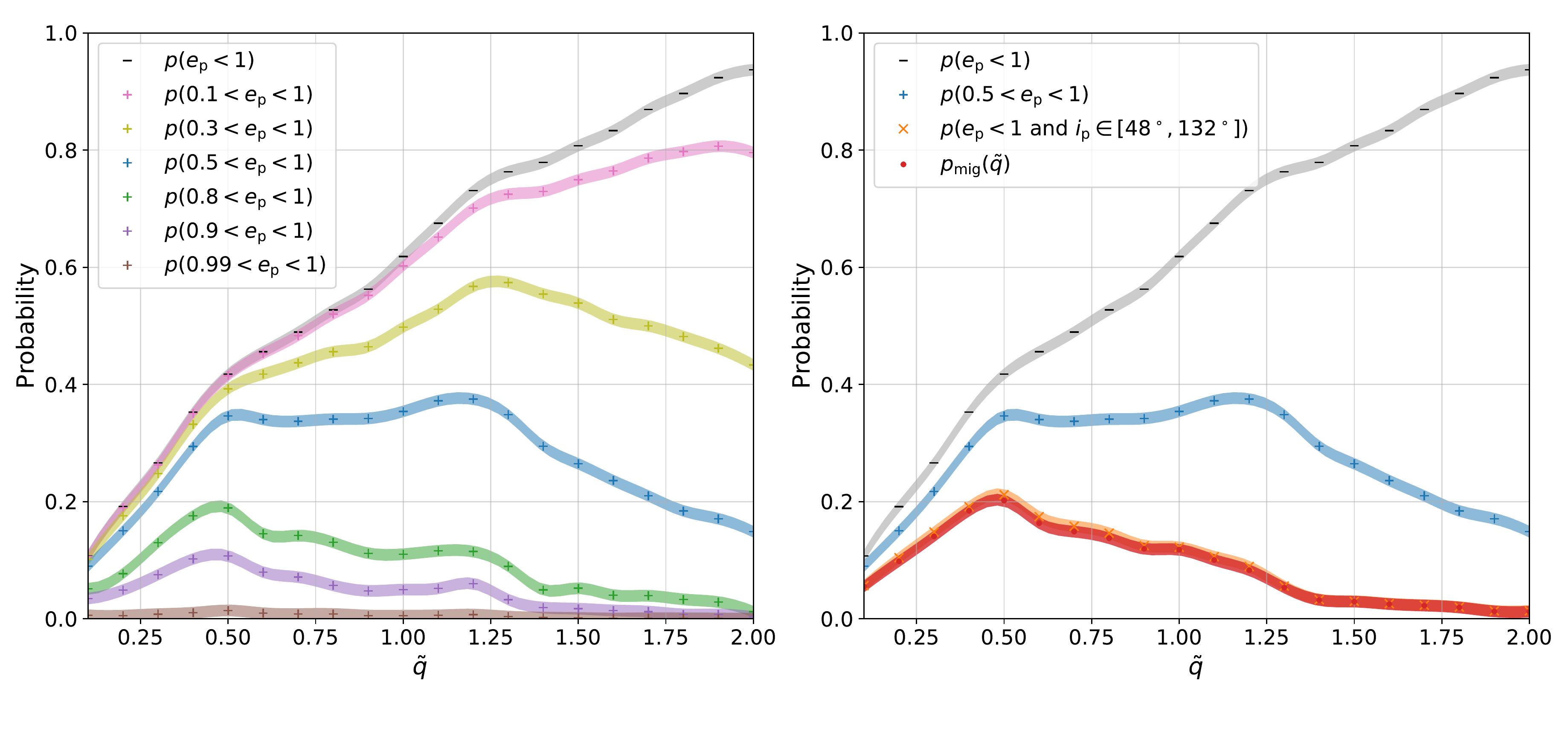}
	\caption{Various probabilities for the outer companion's eccentricity $e_{\rm p}$ and inclination $i_{\rm p}$ to be in critical ranges after a stellar flyby, as functions of the dimensionless distance at the closest approach of the passing star, $\tilde{q} \equiv q/a_{\rm p}$ (where $a_{\rm p}$ is the initial semimajor axis of the companion). The probabilities are computed numerically following the procedure described in Section~\ref{sec:flyby}. The red dots (right panel) show $p_{\rm mig}$, which is the probability for the companion to be in the ``extreme'' ZLK window after a flyby, i.e. the outer companion reaches a sufficiently large inclination and eccentricity to drive the inner planet into high-e migration. This extreme ZLK window is approximately equivalent to $i_{\rm p}\in [48^\circ,132^\circ]$ and $\varepsilon_{\rm oct} > 0.05$, the latter condition corresponds to $0.5 < e_{\rm p} < 1$ for $a_{\rm J}/a_{\rm p} = 1/10$. Note that $p_{\rm mig}$ is almost equal to the probability that the outer companion reaches $i_{\rm p}\in [48^\circ,132^\circ]$ and remains bound ($e_{\rm p} < 1$), indicating that the eccentricity condition is almost always met when the inclination condition is fulfilled.}\label{fig:probaq}
\end{figure*}

In addition to the periastron distance $q$, the outcome of a flyby depends on the inclination $i$ between the flyby orbital plane and the initial orbital plane of the companion, $\omega$, the argument of periastron, and $\lambda_{\rm p}$, the initial orbital phase of the companion. Note that since the initial companion orbit is circular, the outcome of an encounter does not depend on $\Omega$, the longitude of node. In our simulations, we sample the periastron $q$ over a uniform grid from $0.1 a_{\rm p}$ to $2 a_{\rm p}$, and choose
\begin{itemize}
	\item $\omega$ with a flat prior between $0$ and $2 \pi$,
	\item $i$, with a $\sin(i)$ probability distribution, between $0$ and $\pi$,
	\item $\lambda_{\rm p}$ with a flat prior between $0$ and $2 \pi$.
\end{itemize}
For each value of $q$, we sample $20\times20\times20$ angles (for $\omega$, $i$, and $\lambda_{\rm p}$) to determine the distributions of post-flyby orbital parameters of the companion. Several examples of the post-encounter outcomes for three different $\tilde q$ values are shown on Figure~\ref{fig:several_q}. It is clear that flybys with $\tilde{q} =2$ (and larger) produce a negligible number of systems with $i_{\rm p}\gtrsim 40^\circ$. Thus, for our purpose, there is no need to consider encounters with $\tilde{q} > 2$. 

Using results like Figure~\ref{fig:several_q}, we can then compute, for each $q$, the percentage of systems that experience a successful flyby---i.e., the outer companion $m_{\rm p}$ remains bound and is in the eccentricity and inclination window (see Figure~\ref{fig:fig1}) for inducing extreme eccentricity excitation of the inner planet, assuming the orbit of the inner planet ($m_{\rm J}$) unaffected by the flyby (see below). The resulting probability of successful flybys as a function of $q$ is shown in Figure~\ref{fig:probaq} for $a_{\rm J}/a_{\rm p} = 0.1$ (the uncertainty in the probability due to the discrete sampling of the parameter space is of order 1\%). In this case, $\varepsilon_\mathrm{oct}>0.05$ is equivalent to $e_{\rm p}>0.5$. Since a successful flyby would lead to the inner planet to migrate inward, we shall term this probability $p_\mathrm{mig}$. We see from Figure~\ref{fig:probaq} that $p_\mathrm{mig}$ is approximately equal to the probability of producing high inclination: raising the inclination of the outer companion by $48^\circ$ is the hardest constraint to get a successful flyby. This is in line with previous studies finding that the inclination is harder to raise than the eccentricity \citep{li_cross-sections_2015, wang_hot_2020}. In fact, for the entire range of semimajor axis ratios that we study (Equation~\ref{eq:aratio}), most of the cases with high inclinations have $\varepsilon_{\rm oct} > 0.05$, so that the probability of a successful flyby is most constrained by the requirement of significant inclination excitation ($i_{\rm p} \in [48^\circ, 132^\circ]$). 

\begin{figure}
    \centering
    \includegraphics[width=\linewidth]{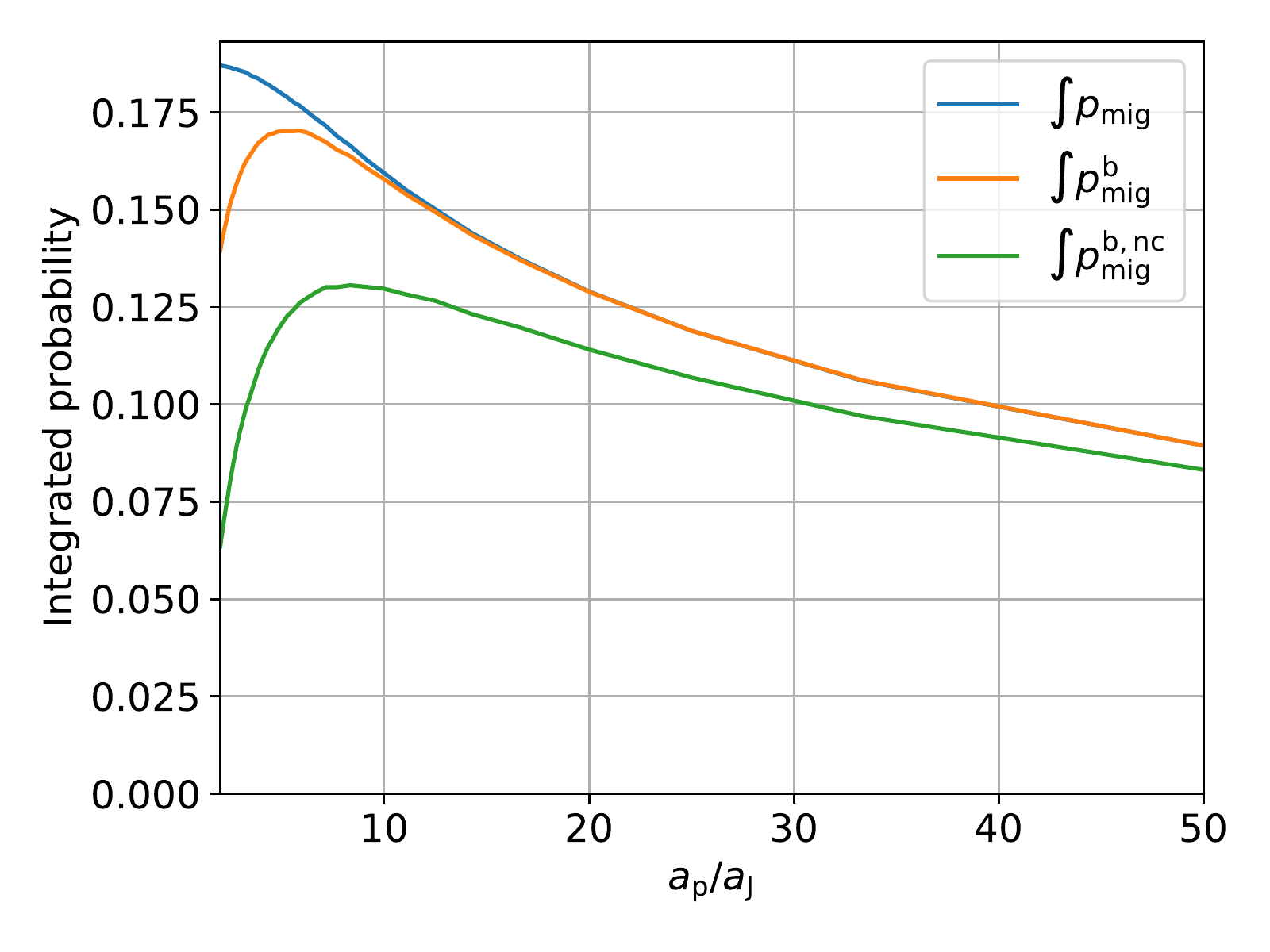}
    \caption{Integrated probability $\int_0^\infty p(\tilde{q}) d\tilde{q}$ for the two-planet system ($m_{\rm J}$ and $m_{\rm p}$) to be in the ``extreme'' ZLK window after a flyby, for different initial semimajor axis ratios $a_{\rm p}/a_{\rm J}$. $\int p_\mathrm{mig}$ is the integrated probability for the outer companion's eccentricity and inclination to be in the critical range (Figure~\ref{fig:probaq}), $\int p_\mathrm{mig}^{\rm b}$ adds the condition that the inner planet should remain bound, and $\int p_\mathrm{mig}^{\rm b, nc}$ additionally requires that the two bodies' orbits cannot cross.}
    \label{fig:pmig_vs_aratio}
\end{figure}

In the above, we have focused on the effect of the flyby on the outer companion. We can similarly use our simulation results to study the effect of the flyby on the inner planet: since the orbits of the two bodies are hierarchical, the perturbations on each from the passing star can be estimated independently. We first require the inner planet to stay bound after the encounter ($p_\mathrm{mig}^{\rm b}$), which can be computed from the initial $a_{\rm p}/a_{\rm J}$ and the probability for a planet to remain bound as a function of $\tilde{q}$ (gray line, Figure~\ref{fig:probaq}). We also require that the periastron of the outer companion be larger than the orbit of the inner one (noncrossing condition $p_\mathrm{mig}^{\rm b, nc}$, $a_{\rm p}(1-e_{\rm p}) > a_{\rm J}$). For $a_{\rm p}/a_{\rm J} < 10$, flybys that strongly perturb the outer companion will also have a non-negligible chance of  disrupting the inner planet. On the other hand, larger $a_{\rm p}/a_{\rm J}$ require a larger $e_{\rm p}$ to reach $\varepsilon_\mathrm{oct} = 0.05$ (see Figure~\ref{fig:fig1}), which will increase the likelihood of orbital crossing. We show the resulting change in the integral of $p_\mathrm{mig}$ over all $\tilde{q}$ (which we will see below, in Equation~\ref{eq:Rmig}, is the relevant quantity for our study) as a function of $a_{\rm p}/a_{\rm J}$ in Figure~\ref{fig:pmig_vs_aratio}. Our estimate of the migration probability will enable us to derive the occurrence rate of successful flybys in Section~\ref{sec:rate}, and to deduce the overall
probability of forming HJs from close stellar encounters.

\section{Occurrence rate}
\label{sec:rate}

In this section, we estimate the occurrence rate of HJs that form in our flyby induced high-e migration scenario (Section~\ref{sec:scenario}). This occurrence rate can be written as
\begin{equation}
	\eta_\mathrm{HJ} = t_\mathrm{cluster} ~ \eta_\mathrm{survival} \int_{10~\mathrm{au}}^{300~\mathrm{au}} \mathrm{d}a_{\rm p}~\dv{\eta_\mathrm{init}}{a_{\rm p}}  ~\mathcal{R}_\mathrm{mig}(a_{\rm p}),  ~ \label{eq:occurrence}
\end{equation}
where $\mathrm{d}{\eta_\mathrm{init}}/\mathrm{d}{a_{\rm p}}$ is the probability distribution function of the initial conditions (two planets, one giant planet around the snow line and one larger companion at $a_{\rm p}$, with angular momentum ratio $l$ less than 0.1), $t_\mathrm{cluster}$ is the lifetime of the birth cluster, $\eta_\mathrm{survival}$ is the probability for the giant planet to survive tidal disruption during high-e migration, and $\mathcal{R}_\mathrm{mig}$ is the rate of close encounters that result in high-e migration. The lower limit of the $a_{\rm p}$ integration ensures dynamical stability, while the upper limit is the consequence of Equation~\eqref{eq:aratio}.

Tidal disruption can limit the efficiency of HJ formation through high-e migration. Using population synthesis models and analytical calculations \citep[e.g.;][]{petrovich_steady-state_2015,anderson_formation_2016,munoz_formation_2016,teyssandier_formation_2019,vick_chaotic_2019}, it has been estimated that most of the giant planets that reach $e_{\rm lim}$ will be destroyed by the tidal forces of their host star. However, there are important uncertainties regarding the fraction of surviving planets, depending on the properties of tidal dissipation. \cite{vick_chaotic_2019} showed that strong dissipation, through a mechanism called chaotic tides, can sometimes save planets otherwise fated for tidal disruption by rapidly decreasing their eccentricities. They estimated that $\eta_\mathrm{survival} \sim 20\%$ of migrating planets could survive as HJs.

Observations provide only limited information on $\mathrm{d}{\eta_\mathrm{init}}/\mathrm{d}{a_{\rm p}}$. Radial velocity surveys suggest an occurrence rate for giant planets between 10\% and 30\%, with a maximum likelihood around 3 au \citep[e.g.,][]{fernandes_hints_2019}. On the other hand, direct imaging surveys point toward an occurrence rate of wide planetary/brown dwarf companions (10--300 au) of around 5-10\% \citep{nielsen_gemini_2019,vigan_sphere_2020}. The correlation between the occurrences of giant planets at a few astronomical units and companions at 10--300 au is not known. Thus, 
\begin{equation}
	\eta_\mathrm{init} \equiv \int_{10~\mathrm{au}}^{300~ \mathrm{au}} \mathrm{d}a_{\rm p}~\dv{\eta_\mathrm{init}}{a_{\rm p}} \sim 0.5-10 \%. \label{eq:etainit}
\end{equation}
The dependency of $\mathrm{d}{\eta_\mathrm{init}}/\mathrm{d}{a_{\rm p}}$ on the semimajor axis $a_{\rm p}$ is unconstrained.


From Section~\ref{sec:flyby}, we can derive the rate of successful flybys, which lead to suitable conditions for HJ formation. This is a function of the cluster stellar density distribution $\mathrm{d}n_\star(v_\infty)$, which is a function of the velocity distribution $f$ of the stellar velocities $v_\infty$. We take $f$ to be a Maxwell--Boltzmann distribution with dispersion $\sigma_\star$:
\begin{eqnarray}
    \mathrm{d}n_\star ={}&& n_\star f(v_\infty) \mathrm{d}v_\infty\\
    f(v_\infty) ={}&& \sqrt{\frac{2}{\pi}} \frac{v_\infty^2}{\sigma_\star^3}   \exp\left(-\frac{v_\infty^2}{2\sigma_\star^2}\right).
\end{eqnarray}
The cluster density $n_\star$ can take a wide range of values, from $10^{-1}$ stars pc$^{-3}$ in the nearby OB associations to $10^{6}$ stars pc$^{-3}$ in the center of globular clusters. Moreover, in an unbound stellar association, which corresponds to most stellar birth environments, the density decreases with time. As a rough estimate, we suppose that our cluster of interest maintains a $10^3$ stars pc$^{-3}$ density for the first $t_\mathrm{cluster} \sim 20$ Myr of its life \citep{pfalzner_early_2013}. We suppose that the flyby rate at later times is negligible due to the much smaller density. The velocity dispersion $\sigma_\star$ is better constrained thanks to observations in nearby associations and cluster \citep[e.g.][]{wright_kinematics_2018}. We adopt $\sigma_\star \approx 1$ km s$^{-1}$. The rate of successful flybys is then:
\begin{equation}
	\mathcal{R}_\mathrm{mig}(a_{\rm p}) = \int_{0}^\infty \mathrm{d}n_\star(v_\infty) v_\infty  \int_{0}^{+\infty}   \pi  ~\mathrm{d}b^2(q, v_\infty)~ p_\mathrm{mig}(\tilde{q}),\label{eq:Rmig_1}
\end{equation}
where $p_\mathrm{mig}$ is depicted in Figure~\ref{fig:probaq}. The impact parameter $b(q, v_\infty)$ associated with an hyperbolic trajectory of periastron $q$ and velocity at infinity $v_\infty$ is
\begin{equation}
	b^2 = q^2\left(1+\frac{2G M_\mathrm{tot}}{q v_\infty^2}\right)  \approx \frac{2G M_\mathrm{tot}q}{v_\infty^2}.
\end{equation}
Equation~\eqref{eq:Rmig_1} then becomes
\begin{equation}
	\mathcal{R}_\mathrm{mig} = \mathcal{R}_\mathrm{close} \int_{0}^{+\infty} \mathrm{d}{\tilde{q}}~ p_\mathrm{mig}(\tilde{q})\label{eq:Rmig}
\end{equation}
where
\begin{eqnarray}
	\mathcal{R}_\mathrm{close} \equiv{}&& 2\pi a_{\rm p} G M_\mathrm{tot} n_\star \int_{0}^\infty \frac{dv_\infty}{v_\infty} f(v_\infty)\nonumber \\
	{}={}&& \frac{2 \sqrt{2\pi} a_{\rm p} G M_\mathrm{tot} n_\star}{\sigma_\star}\nonumber\\
	\approx{}&&  10 \left(\frac{n_\star}{10^3~ \mathrm{pc}^{-3}}\right) \left(\frac{M_\mathrm{tot}}{2~\mathrm{M_\odot}}\right)\left(\frac{a_{\rm p}}{50~\mathrm{au}}\right)\nonumber\\
	&&{}\times\left(\frac{\sigma_\star}{1~\mathrm{km/s}}\right)^{-1} \mathrm{Gyr}^{-1}\label{eq:Rclose}
\end{eqnarray}
is the rate of close encounters with periastron below $a_{\rm p}$. The integral of $p_\mathrm{mig}$ over all $\tilde{q}$ in Equation~\eqref{eq:Rmig} can be computed numerically using the results from Section~\ref{sec:flyby} (see Figure~\ref{fig:pmig_vs_aratio}). Overall, the integrated probability to induce high-eccentricity migration through ZLK is comprised between $7$\% and $13$\% for our range of $a_{\rm p}/a_{\rm J}$. Thus, a reasonable estimate is
\begin{equation}
	\int_{0}^{+\infty} \mathrm{d}{\tilde{q}}~ p_\mathrm{mig}^{\rm b,nc}(\tilde{q}) \approx 0.10
\end{equation}



Combining Eqs.~\eqref{eq:occurrence}, \eqref{eq:etainit},\eqref{eq:Rmig}, and \eqref{eq:Rclose}, we have
\begin{eqnarray}
	\eta_\mathrm{HJ} &{}\approx{}& 0.05\% ~\left(\frac{M_\mathrm{tot}}{2~\mathrm{M_\odot}}\right)\left(\frac{\langle a_{\rm p} \rangle}{50~\mathrm{au}}\right)\nonumber\\&& \times \left(\frac{\eta_\mathrm{init}}{10\%}\right) \left(\frac{\mathcal{R}_\mathrm{mig}/\mathcal{R}_\mathrm{close}}{10\%}\right) \left(\frac{\eta_\mathrm{survival}}{20\%}\right) \nonumber\\&& \times  \left(\frac{n_\star}{10^3~ \mathrm{pc}^{-3}}\right)\left(\frac{t_\mathrm{cluster}}{20~ \mathrm{Myr}}\right) \left(\frac{\sigma_\star}{1~\mathrm{km/s}}\right)^{-1},  \label{eq:etaHJ}
\end{eqnarray}
where
\begin{equation}
	\langle a_{\rm p} \rangle \equiv \frac{1}{\eta_\mathrm{init}}\int_{10~\mathrm{au}}^{300~ \mathrm{au}} \mathrm{d}a_{\rm p}~\dv{\eta_\mathrm{init}}{a_{\rm p}} a_{\rm p}.
\end{equation}
Equation~\eqref{eq:etaHJ} gives an estimate for the occurrence rate of HJs produced in our scenario, and illustrates its dependence on various uncertain parameters. Major uncertainties include $\eta_\mathrm{init}$, the occurrence rate of the initial two-planet systems, and the density and lifetime of the birth clusters, the latter two may vary by orders of magnitude.

The observed occurrence rate of HJs around solar-type stars is 0.5--1\% \citep[e.g.][]{dawson_origins_2018}. With the adopted fiducial parameters in Equation~\eqref{eq:etaHJ}, this formation channel can account for 5--10\% of the observed HJ population. But with more optimistic $n_\star$, $t_\mathrm{cluster}$, and $\eta_\mathrm{init}$ values, our estimated $\eta_\mathrm{HJ}$ can be compatible with the observed value.

\section{Summary and Discussion}
\label{sec:conclusion}

\subsection{Summary}

In this paper, we have studied a new scenario for the formation of HJs following a close stellar encounter---we call it ``flyby Induced High-e Migration''. This could account for the recently observed correlation between the occurrence of HJs and stellar overdensities \citep{winter_stellar_2020}. In this scenario, we suppose that stellar flybys could excite the eccentricity and inclination of an outer companion (giant planet, brown dwarf, or low mass star, at $a_{\rm p}\sim 10$--300 au), which would trigger the high-e migration of an inner cold Jupiter (initially at a few astronomical units). High-e migration requires the outer body to be in a suitable window of inclination and eccentricity to induce extreme ZLK oscillations of the inner planet (Section~\ref{sec:kozai}). We carry out simulations of the secular evolution of two-planet systems to determine the required ``migration'' window driven by the octupole ZLK effect, extending previous test-particle results. Through extensive $N$-body numerical experiments, we find that for a stellar flyby to have a significant impact on the two-planet system (more specifically, to produce sufficient inclination and eccentricity in the outer companion, and thereby to trigger high-e migration of the inner planet), its closest approach should be comparable to the semimajor axis of the outer body. We then estimate the rate of such ``successful'' encounters and the likelihood that they will lead to the formation of HJs, taking into account geometric and stellar density parameters, as well as tidal disruption of migrating giant planets and the probability to have a suitable planetary system in the first place. Equation~\eqref{eq:etaHJ} gives the resulting occurrence rate of HJs produced in this scenario and its dependence on various parameters. Although the estimated occurrence rate relies on a few poorly constrained parameters, our analysis suggests that this HJ formation channel requires the birth cluster to retain a high stellar density for more than 20 Myr in order to account for a significant fraction of the observed HJ population.


\subsection{Main Uncertainties}

No further analysis or numerical simulations can significantly refine our estimate (Equation~\eqref{eq:etaHJ}) until a better understanding on the properties of the typical stellar birth cluster is obtained, in particular the stellar density $n_\star$ as a function of time.

Furthermore, the correlation between cold Jupiters and more distant planets or brown dwarfs (or low-mass stars) is currently unknown. So the occurrence rate $\eta_\mathrm{init}$ of the initial systems is not constrained. In fact, if we consider the outer body ($m_{\rm p}$) to be a low-mass star, $\eta_\mathrm{init}$ could be much larger than 10\% assumed in Equation~\eqref{eq:etaHJ}. Our knowledge should improve with the next generation of exoplanet and substellar imaging surveys.

\subsection{Comparison with Other High-e Migration Paths}

In the following we consider several other possible mechanisms for HJ formation. All of these are less promising for explaining the correlation between HJs and stellar overdensities reported by \cite{winter_stellar_2020}.

\subsubsection{Planet--Planet Scattering}

Planet--planet scattering is another possible path to create high-eccentricity orbits, which could then lead to high-e migration of giant planets. \cite{wang_hot_2020} determined that it could account for a significant formation of HJs in their simulations involving stellar flybys. This mechanism does not require raising the inclination of the outer planet to enter the ZLK regime, but only exciting the eccentricity enough to prompt a close encounter. How high the eccentricity needs to be raised depends on the semimajor axis ratio $a_{\rm J}/a_{\rm p}$. For $a_{\rm J}/a_{\rm p}=0.5$, the required eccentricity for close planet-planet encounter is about 0.5, and we found from our flyby simulations that the probability can be $\sim 40$\% for $\tilde q \lesssim 2$ (see Figure~\ref{fig:probaq}). For $a_{\rm J}/a_{\rm p}=0.1$, the required eccentricity is 0.9, and our simulations suggest a much smaller probability ($\lesssim 10$\%; see Figure~\ref{fig:probaq}). Note that as a general rule, we found that a flyby more readily raises the eccentricity than the inclination of the planet.


However, the probability that planet-planet scattering leads to HJ formation is likely much smaller. The eccentricity of the inner planet must increase sufficiently to enter the effective range of tidal dissipation, while remaining bound to the system. The outcome of an encounter between the planets varies with the conditions of the encounter and with the semimajor axis, divided between ejection, planet-planet merger, collision with the star, and survival as an HJ. This problem has recently been studied by \cite{li_giant_2021} for initially quasi-circular orbits. In their simulations, the probability that a planet undergoes a close encounter with the host star and becomes an HJ is less than 1\%. Although in our case one of the planets has a highly eccentric orbit, the order of magnitude is likely similar. Thus, following a stellar flyby, planet-planet scattering seems unlikely to be more common than ZLK excitation to trigger high-e migration.

\subsubsection{Stellar Companion}

The probability for a solar-type star to have a stellar companion at separations 100--1000 au is close to 20\% in the field \citep{raghavan_survey_2010}. Assuming a flat eccentricity and an ``isotropic'' inclination distributions, a significant fraction of binary companions could induce ZLK migration of giant planets. This
HJ formation channel has been well studied in the literature \citep[see][and references therein]{petrovich_steady-state_2015,anderson_formation_2016,munoz_formation_2016,vick_chaotic_2019}. The predicted HJ occurrence rate is of order 0.1-0.2\%, with one of the main uncertainties being tidal disruption of the migrating planet. This formation channel seems more efficient at forming HJs because of the higher occurrence rate of stellar companions at large separation and, most importantly, because it does not require a flyby.

However, this formation channel struggles to account for the observed correlation between HJs and stellar overdensities \citep{winter_stellar_2020}. While the dependence of the stellar multiplicity on the stellar density is not well constrained, several surveys in nearby clusters seems to indicate that overdensity actually decreases stellar multiplicity \citep{king_testing_2012,marks_inverse_2012}.

\subsubsection{Tides from the Stellar Cluster}

Stellar clusters create a global gravitational potential, which can trigger eccentricity and inclination variations on wide companions \citep{hamilton_secular_2019-1,hamilton_secular_2019}. Similar to the ZLK mechanism, here the ``perturber'' is the global potential of the cluster. For a planetary system located at a distance $R_{\rm c}$ from the center of the cluster, the eccentricity excitation timescale is of order 
\begin{eqnarray}
    T&&{}\sim \frac{R_{\rm c}^3}{GM_{\rm c}} n_{\rm p} \sim \frac{3 n_{\rm p}}{4\pi G\bar\rho_\star}\nonumber\\
    &&{}\sim 1~\mathrm{Gyr} \left(\frac{10^3~\mo/\pc^3}{\bar\rho_\star}\right) \left(\frac{50~\mathrm{au}}{a_{\rm p}}\right)^\frac{3}{2}  \left(\frac{M_1}{1~\mo}\right)^\frac{1}{2}, \label{eq:timescale}
\end{eqnarray}
where $n_{\rm p}$ and $a_{\rm p}$ are the (initial) mean motion and semimajor axis of the planet, $M_{\rm c}$ is the enclosed cluster mass at $R_{\rm c}$, and $\bar\rho_\star$ is the mean density of the cluster. This timescale is much larger than the lifetime of the cluster. Moreover, at such a density, about 20 close flybys are expected to occur within 1 Gyr (Equation~\ref{eq:Rclose}), and their influence will be dominant on the dynamics of the planet. 


\section*{Acknowledgements}

This work has been supported in part by the NSF grant AST-17152 and NASA grant 80NSSC19K0444. Y.S. is supported by the NASA FINESST grant 19-ASTRO19-0041. We made use of the \textsc{python} libraries \textsc{NumPy} \citep{harrisArrayProgrammingNumPy2020}, \textsc{SciPy} \citep{virtanenSciPyFundamentalAlgorithms2020}, and \textsc{PyQt-Fit}, and the figures
were made with \textsc{Matplotlib} \citep{hunterMatplotlib2DGraphics2007}.

\bibliography{Biblio}{}
\bibliographystyle{aasjournal}

\end{document}